\documentclass[a4,11pt]{cip-submit-2019}  
\usepackage{color,soul}
\usepackage{afterpage}

\usepackage{wrapfig}
\usepackage{wasysym}
\usepackage{enumitem}
\usepackage{adjustbox}
\usepackage{ragged2e}
\usepackage[svgnames,table]{xcolor}
\usepackage{tikz}
\usepackage{longtable}
\usepackage{changepage}
\usepackage{setspace}
\usepackage{hhline}
\usepackage{multicol}
\usepackage{tabto}
\usepackage{float}
\usepackage{multirow}
\usepackage{makecell}
\usepackage{mathptmx}
\DeclareSymbolFont{epsilon}{OML}{cmm}{m}{it}
\DeclareMathSymbol{\epsilon}{\mathord}{epsilon}{"0F}
\def\DD{D\kern-.7em\raise0.25ex\hbox{\char '55}\kern.33em}
\usepackage{hyperref}
\hypersetup{
	colorlinks=false,
	pdfborder={0 0 0}
}
\usepackage{graphicx,subfigure,wrapfig,epstopdf}
\def\Mr{\uppercase}
\usepackage{graphics}
\usepackage{wasysym}%\permil 
\usepackage{amsmath,amsxtra,amssymb,latexsym,amscd,color,cite,bm}
\usepackage[mathscr]{eucal}
\usepackage{array}
\usepackage{multirow}
\usepackage{cite} 
\def\doi#1{\href{http://dx.doi.org/#1}{doi:#1}}
\def\titles#1{\title{\large\bf\noindent #1}}
\def\authors#1{\author{\begin{flushleft}{#1}\end{flushleft}}}
\def\authord#1#2{\indent\Mr{#1}$^{#2}$}
\def\addressed#1#2{\\[1mm]\textit{$\!\!\!^{#1}$\indent#2}}
\def\Email{$^{\dagger}$}
\def\email#1{\bigskip\href{mailto:#1}{\!\!\Email\textit{E-mail:}~{#1}}\\[3mm]}
\def\PublicationInformation#1#2#3#4{\\[4mm]\href{mailto:#1}{\!\!\Email\textit{E-mail:}~{#1}}\\[3mm]
	\textit{\indent Received #2}\\[1mm]
	\textit{Accepted for publication~#3}\\[1mm]
	\textit{Published~#4}}
\def\Keywords#1{\\[.2cm] \textnormal{Keywords:~{#1}}.} 

\def\and{$\text{\tiny AND }$}

\def\Classification#1{\\[.2cm] \textnormal{Classification numbers:~{#1}.}}  
\newcommand{\sout}[1]{\unskip}
\def\ed{
	\bibliographystyle{cip-sty-2019}
	\bibliography{references-database-name}
\end{document}
}
\begin{document}
%PAGE_NUMBER%%%%%%%%%%%%%%
%	\Volume{??}\Number{??}	\ArticleId{????}
	\Year{2022}
	\Page{1}\Endpage{11}
%%%%%%%%%%%%%%%%%%%%%%%%%%%%%%%%%%%%%%%%%%%%%
	\titles{Heat conductance oscillations in two weakly connected charge Kondo circuits}
	%%%%%%%%%%%%%%%%%%%%%%%%%%%%%%%%%%%%%%%%%%%%%%%%%
	%%%%%AUTHORS AND ADDRESSES 
	%%%%%%%%%%%%%%%%%%%%%%%%%%%%%%%%%%%%%%%%%%%%%%%%%
	\authors{
	\authord{T. K. T. Nguyen}{1}\Email and \authord{M. N. Kiselev}{2} 
	%%%%%%%%%%%%%%%%%%%
	\newline
	%%%%%%%%%%%%%%
	\addressed{1}{Institute of Physics, Vietnam Academy of Science and Technology,\\
		10 Dao Tan, Ba Dinh, Hanoi, Vietnam}
	\addressed{2}{The Abdus Salam International Centre for Theoretical Physics,\\
		 Strada Costiera 11, I-34151, Trieste, Italy}
%		\email{nkthanh@iop.vast.ac.vn}%Corressponding author
%		\received{\today}
%		\accepted{DD MM YYYY}
%\PublicationInformation{Email}{received date}{Accepted date}{Published date}
\PublicationInformation{nkthanh@iop.vast.vn}{\today}{??}{??}
	}
	\maketitle
%%%%%%%%%%%%%%%%%%%%%%%%%%%%%%%%%%%%%%%%%%%%%%%%%%%%%%%%%%%%%%%%%%%%%%%
	\markboth{SHORTEN NAME OF AUTHORS ...}{TITLE OF THE ARTICLE...}
%%%%%%%%%%%%%%%%%%%%%%%%%%%%%%%%%%%%%%%%%%%%%%%%%%%%%%%%%%%%%%%%%%%%%%%

\begin{abstract}
We revisit a model describing the Seebeck effect on a weak link between two charge Kondo circuits, which has been proposed in the {\it [Phys. Rev. B \textbf{97}, 085403 (2018)]}. We calculate the thermoelectric coefficients in the perturbation theory assuming smallness of the reflection amplitudes of the quantum point contacts. We focus on the linear response equations for the heat conductance in three different scenarios as: Fermi liquid vs Fermi liquid, Fermi liquid vs non-Fermi liquid, non-Fermi liquid vs non-Fermi liquid. The oscillations of the heat conductance  as a function of the gate voltage of each quantum dot are analyzed in both Fermi liquid and non-Fermi liquid regimes. We discuss possible experimental realizations of the model to observe the signatures of the  non-Fermi liquid behavior in the heat conductance measurements.
\Classification{73.23.Hk, 73.50.Lw, 72.15.Qm, 73.21.La}
\end{abstract}

\section{\Mr{Introduction}}

Controlling the thermal properties of electronic devices can support
technologies based on heat management. A significant amount of work
has already been performed to understand heat conduction in nanodevices
\cite{Zlatic_book,review}. It is known that the role of quantum electron
transport in many nanostructures affects not only the charge and spin
transport phenomena but also the heat transport mechanism. The study
of quantum transport, especially the thermoelectric transport in nano-structured
devices at very low temperature, is thus an important and rapidly
developing topic in recent years. Among a large variety of available
nano-devices, {\color{black} quantum dots (QDs) play an} important
and significant role \cite{BN_book}. {\color{black} On the one
hand, QDs are highly controllable and finely adjustable by external
fields. On the other hand, QDs being part of the quantum circuits
demonstrate pronounced effects of the electron-electron interactions,
resonance scattering and quantum interference observable in the quantum
transport experiments.}

{\color{black} One of the remarkable many-body effects in QD devices
is the Kondo effect \cite{Kondo_1964}. A single electron transistor
QD device in the Kondo regime shows different universal behaviors
at different energies \cite{Hewson_book}. The conventional single
impurity $S=1/2$ single channel Kondo model (1CK) is characterized
by a unique energy scale known as the Kondo temperature $T_{K}$ determining
the universal behavior of the quantum thermodynamic and transport
observables both at $T>T_{K}$ and $T<T_{K}$. The impurity spin of
1CK becomes completely screened by the mobile electrons at $T\to0$.
At energies greater compared to the Kondo temperature $T_{K}$ ($T_{K}$
plays the role of the Fermi energy for 1CK), the system properties
can be accessed through the perturbation theory approach. This regime
is known as a weak coupling Kondo regime \cite{Kondo_1964,Hewson_book}.
The behavior of $M$-orbital spin-$S$ Kondo model at the energies
lower compared to the Kondo temperature $T_{K}$ depends on the way
the mobile electrons screen the impurity spin. For instance, for the
full screened case in which $M=2S$ the system at $T<T_{K}$ is in
the strong coupling regime, corresponding to the strong coupling fixed
point in the renormalization group (RG) flow diagram \cite{Hewson_book}.
This regime is coherent and the system is mentioned having Fermi-liquid
(FL) properties. It happens the same when the system is underscreened
($M<2S$). However, for the overscreened case ($M>2S$) which is characterized
by intermediate unstable coupling fixed point, the system is potentially
having the non-Fermi liquid (NFL) characteristics \cite{Hewson_book,Nozieres_Blandin_1980}.
Recent studies of the multi-channel Kondo physics in QD devices have
focused on the $T<T_{K}$ limit \cite{flensberg,matveev,furusakimatveevprb,andreevmatveev,LeHur2002,thanh2010,nk2015,thanh2018,thanhprl}.
It is found that it is hard to achieve the strong NFL regime at very
low temperature, namely $T\ll T^{*}<T_{K}$, where $T^{*}$ is related
to the parameter of the perturbative expansion $|r|$ as $T^{*}=|r|^{2}T_{K}$.
The system has a tendency to fall into the FL regime associated with
stable FL fixed point \cite{thanh2010}. Nevertheless, at higher temperature
regimes, $T^{*}<T<T_{K}$, the fingerprints of the weak NFL behavior
can be observed \cite{thanh2010}.}

While the conventional Kondo phenomenon is attributed to a spin degree
of freedom of the quantum impurity, the unconventional charge Kondo
effect deals with an iso-spin implementation of the charge quantization.
Recently, the breakthrough experiments \cite{pierre2,pierre3} have
been successful in implementing a setup that consists of a large metallic
QD electrically connected to two-dimensional electron gas (2DEG) electrodes
through quantum point contacts (QPCs). A strong magnetic field is
applied perpendicularly to the 2DEG plane. The 2DEG is in the Integer
Quantum Hall (IQH) regime. The QPCs are fine-tuned to satisfy the
condition that only one chiral edge current is partially transmitted
across the QPCs. The logic behind the mapping of IQH setup to a multi-channel
Kondo (MCK) problem has been explained in Refs. \cite{thanh2018,thanhprl}.
Namely, if we assign the ``iso-spin up'' to the electrons inside
the QD and the ``iso-spin down'' to the electrons outside the QD,
the charge iso-spins flip at QPCs as the backscattering transfers
the ``moving in-'' QD electrons to ``moving out-'' QD electrons
and vice versa. The number of QPCs is equivalent to the number of
orbital channels in the conventional $S=1/2$ Kondo problem. Therefore,
these experimental setups allow us to investigate the properties of
one- or multi-channel Kondo systems characterized by FL or NFL behaviors
correspondingly. These experiments mark an important step in the study
of the MCK problems. Indeed, fairly recently, another experimental
study \cite{two_islands} has successfully implemented a tunable nanoelectronic
circuit comprising two coupled hybrid metallic-semiconductor islands,
combining the strengths of the two types of materials, which can demonstrate
a potential for scalability and a novel quantum critical point.

In the recent years, the thermoelectric transport through QD systems
has attracted attentions of both theorists \cite{furusakimatveevprb,andreevmatveev,Beenakke,Turek_MAtveev}
and experimentalists \cite{Staring_europhysics1993,Dzurak_exp,kondo_th_exp,Kondo_exp_GG2007}.
{\color{black} QDs have advanced applications in thermoelectricity
and microelectronics \cite{Zlatic_book,apply_thermoelectricity,apply_TE}
and can be used as the} tools for a better understanding of closely
correlated systems \cite{correlation_Zlatic_Hewson,correlation_Kim}.
Among all the thermoelectric coefficients, the thermopower is the
most interesting object due to its high sensitivity to the particle-hole
asymmetry of the system. The thermoelectric measurements allow to
investigate the effects related to the hole-particle asymmetry and
provide information on low-energy excitations in the system \cite{andreevmatveev,Vavilov_Stone,KK2019}.
These properties of thermopower open a possibility for capturing the
FL -- NFL crossover by accessing the NFL mode \cite{thanh2010,thanh2018,KK2019}.
Lately, the extended studies \cite{karki1,karki2,Pavlov_Kiselev_2021}
have investigated heat conductance in order to better demonstrate
the FL and NFL pictures in QD systems. Moreover, understanding thermoelectric
properties of a system promotes the study of entropy \cite{entropy1,entropy2,entropy3,entropy4}.

\begin{figure}
\centering \includegraphics[width=0.7\columnwidth]{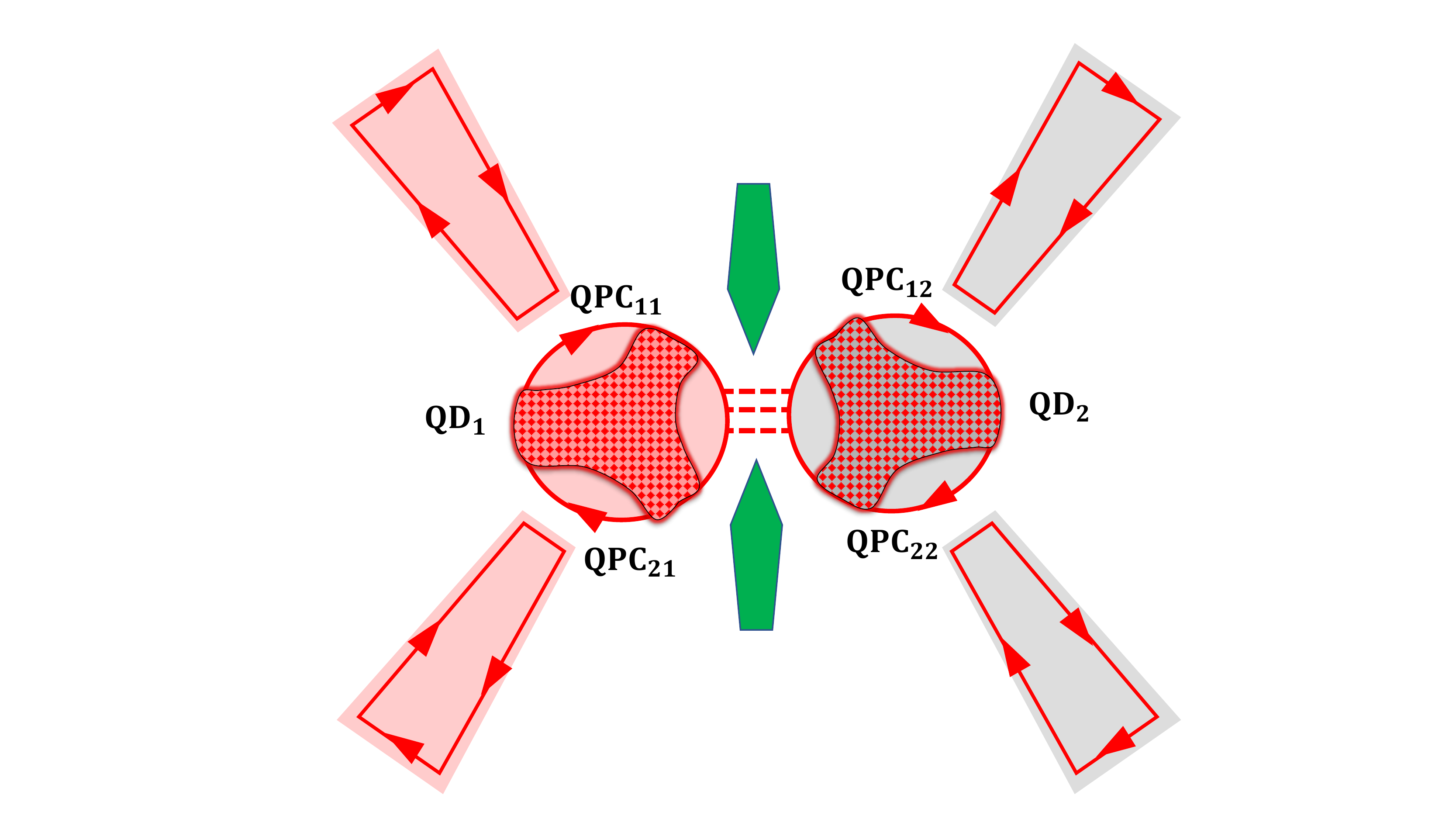} \caption{(Color online) Schematic of two charge Kondo site setup. Each site
consists of a large metallic quantum dot (QD) (the patch) connected
to a two dimensional electron gas (2DEG) (the region inside the circle)
and electrode (the trapezoid) through quantum point contacts (QPCs).
Controlling the transparencies of the QPCs provides a weak coupling
between a) two Fermi liquids; b) a Fermi liquid and a non-Fermi liquid;
c) two non-Fermi liquids. Two QDs are connected through the tunnel
barrier (red dashed lines) which is controlled by a voltage (the green
patches). The pink color stands for the higher temperature $T+\Delta T$
compared to the reference temperature $T$ of the gray color. The
currents flowing along the chiral spin edge channel are denoted by
red lines with arrows. They are partially transmitted through the
almost transparent QPCs. }
\label{f1} 
\end{figure}

In this work, we revisited the setup which has been proposed in 2018
\cite{thanh2018}. The generalization of the ideas of Flensberg-Matveev-Furusaki
(FMF) theory \cite{flensberg,matveev,furusakimatveevprb} is adopted
to describe the IQH charge Kondo nanodevices \cite{pierre2,pierre3}.
The design for the quantum-dot--quantum-point contact (QD-QPC) devices
for investigation of weakly coupled Fermi and non-Fermi liquid states
is shown in Fig \ref{f1} which can be one of three cases: a) two
Fermi liquids; b) a Fermi liquid and a non-Fermi liquid; c) two non-Fermi
liquids. We compute perturbatively the thermoelectric coefficients
and concentrate on the heat conductance. Discussing the behavior of
the heat conductance as a function of temperature and gate voltages
in all three cases we find that the FL behavior is more prominent
than the NFL one.

The paper is organized as follows. We describe the theoretical model
for observing the FL and NFL behavior in Sec. II. General equations
for the thermoelectric coefficients are presented in Sec. III. The
main results are shown in Sec. IV. We conclude our work in Sec. V.

\section{\Mr{Model}}

We consider a setup (see Fig. \ref{f1}) consisting of two QD-QPC
structures weakly coupled through the tunnel barrier between two QDs.
Each large metallic QD with a continuous spectrum is electrically
connected to a two-dimensional electron gas (2DEG) denoted by pink
and gray areas inside circles and further connected to a large electrode
through several QPCs. The 2DEG is in the IQH regime at filling factor
$\nu=2$ by applying a strong quantizing magnetic field perpendicularly
to it. The QPCs are fine-tuned to achieve a regime where the current
flows along the outer spin-polarized edge channel (shown by red color
on Fig. \ref{f1}) is partially transmitted across QPCs. The inner
channel (not shown on Fig. \ref{f1}) is fully reflected and can be
ignored. The logic behind the mapping of IQH setup to a single/multichannel
Kondo problem has been explained in Refs. \cite{thanh2018,thanhprl},
so each QD-QPC structure is a single or multi-channel charge Kondo
simulator. At the weak link between two QDs, temperature drop happens,
namely, the pink color stands for the higher temperature $T+\Delta T$
compared to the reference temperature $T$ of the gray one.

The spinless Hamiltonian describing the two QD-QPC structure coupled
weakly at the center (Fig. \ref{f1}) has the form $H$$=$$H_{1}+H_{T}+H_{2}$,
where 
\begin{eqnarray}
H_{T}=(td_{1}^{\dagger}d_{2}+\text{h.c.}).\label{h0}
\end{eqnarray}
describes the tunneling between two dots, {\color{black}$d_{j}$
is the annihilation operator of an electron in the dot $j,$ ($j=1,2$)
at the position of the weak link. } Each $\text{QD}_{j}$ is coupled
strongly to the leads through $\text{QPC}_{\alpha j}$ ($\alpha=1,2$)
so that the whole part {[}$\text{(QD-QPC)}_{j}$ structure{]} is described
by Hamiltonian $H_{j}$$=$$H_{0j}+H_{Cj}+H_{sj}$. {\color{black}
The Hamiltonian $H_{0,j}$ stands for the free part representing two
copies of free one-dimensional electrons in $\text{QPC}_{\alpha j}$
($\alpha j=11,21,12,22$, see Fig. \ref{f1}) and corresponding patch
areas between $\text{QPC}_{\alpha j}$ and $\text{QD}_{j}$ of the
$\text{(QD-QPC)}_{j}$: 
\begin{equation}
H_{0,j}=\sum_{\alpha=1,2}-iv_{F}\int_{-\infty}^{\infty}dx\left[\psi_{\uparrow,\alpha,j}^{\dagger}\partial_{x}\psi_{\uparrow,\alpha,j}-\psi_{\downarrow,\alpha,j}^{\dagger}\partial_{x}\psi_{\downarrow,\alpha,j}\right].
\end{equation}
} {\color{black} Here we define $\psi_{\uparrow,\alpha,j}$ ($\psi_{\downarrow,\alpha,j}$)
the operator describing one-dimensional fermions which are inside
(outside) the $\text{QD}_{j}$ in the $\text{(QD-QPC)}_{j}$, $v_{F}$
is a Fermi velocity. We adopt the units $\hbar=c=k_{B}=1$ in this
paper.}

The QDs are assumed in the Coulomb blockade regime which is demonstrated
by the Hamiltonian $H_{C}$: {\color{black} 
\begin{equation}
H_{C,j}=E_{C,j}\left[\hat{n}_{t,j}+\hat{n}_{QPCs,j}-N_{j}(V_{g,j})\right]^{2},\label{charging-0}
\end{equation}
with $E_{C,j}$ is charging energy, $\hat{n}_{t,j}$ and $\hat{n}_{QPCs,j}=\int_{0}^{\infty}\sum_{\alpha}\psi_{\alpha,j}^{\dagger}(x)\psi_{\alpha,j}(x)\,dx$
\cite{aleiner_glazman} are the operator of the number of electrons
that entered the QD$\!_{j}$ through the weak tunnel and the QPCs
correspondingly, and $N_{j}$ is a dimensionless parameter which is
proportional to the gate voltage $V_{g,j}$.} The Hamiltonian $H_{s,j}$
describes the backward scattering at the $\text{QPC}_{\alpha j}$
on the side $j$, which is controlled by {\color{black} a short-range
isospin-flip} voltage $V_{\alpha j}\left(x\right)$: {\color{black}
\begin{align}
H_{s,j} & =\sum_{\alpha=1,2}\int_{-\infty}^{\infty}dx\left[\psi_{\uparrow,\alpha,j}^{\dagger}\left(x\right)V_{\alpha j}\left(x\right)\psi_{\downarrow,\alpha,j}\left(x\right)e^{-i2k_{F}x}+h.c.\right].
\end{align}
In the spirit of FMF theory \cite{flensberg,matveev,furusakimatveevprb},
we replace $d_{j}^{\dagger}d_{j}=\sum_{\alpha}\psi_{\uparrow,\alpha,j}^{\dagger}(-\infty)\psi_{\uparrow,\alpha,j}(-\infty)F^{\dagger}F$,
where $F$ is the operator lowering $\hat{n}_{t,j}$ by unity. The
operator $F^{\dagger}(t)$ increases $n_{t,j}$ from $0$ to $1$
at time $t$, while $F(t)$ decreases $n_{t,j}$ back to $0$ from
$1$. In the bosonic represention, the fermionic operator is related
to the bosonic operator as: $\psi_{(\uparrow/\downarrow),\alpha,j}(x)\sim e^{i\phi_{\alpha,j}(x)}$
with} $\phi_{\alpha,j}$ is a bosonization displacement operator
describing transport through the $\text{QPC}_{\alpha j}$ of the $\text{(QD-QPC)}_{j}$,
with a scatterer at $x$$=$$0$. We can rewrite the $H_{C,j}$ ,
$H_{0,j}$ and $H_{s,j}$ in the bosonic language as follows. 
\begin{eqnarray}
H_{C,j}=E_{C,j}\left[\hat{n_{j}}+\frac{1}{\pi}\sum_{\alpha}\phi_{\alpha,j}(0)-N_{j}(V_{g,j})\right]^{2},\label{charging}
\end{eqnarray}
\begin{equation}
H_{0,j}=\sum_{\alpha}\frac{v_{F}}{2\pi}\!\!\!\int_{-\infty}^{\infty}\!\!\!\!\!\!\left\{ [\Pi_{\alpha,j}(x)]^{2}+[\partial_{x}\phi_{\alpha,j}(x)]^{2}\right\} dx,\label{hr}
\end{equation}
\begin{equation}
H_{s,j}=-\frac{D}{\pi}\sum_{\alpha}|r_{\alpha j}|\cos[2\phi_{\alpha,j}(0)],
\end{equation}
{\color{black} where $\Pi_{\alpha,j}$ is the conjugated momentum
$[\phi_{\alpha,j}(x),\Pi_{\alpha',j}(x')]=i\pi\,\delta(x-x')\delta_{\alpha\alpha'}$,
$D$ is a bandwidth, $r_{\alpha j}=-iV_{\alpha j}(2k_{F})/v_{F}$
is the reflection amplitude of the $\text{QPC}_{\alpha j}$. }

\section{\Mr{Heat current and heat conductance}}

We sketch the derivation of the electric and heat currents: 
\begin{align}
I_{e} & =-2\pi e|t|^{2}\int_{-\infty}^{\infty}d\epsilon\nu_{1}(\epsilon)\nu_{2}(\epsilon)\left[f_{1}(\epsilon)-f_{2}(\epsilon)\right],\\
I_{h} & =2\pi|t|^{2}\int_{-\infty}^{\infty}d\epsilon\epsilon\nu_{1}(\epsilon)\nu_{2}(\epsilon)\left[f_{1}(\epsilon)-f_{2}(\epsilon)\right],
\end{align}
with $|t|$ is a modulus of the tunnel matrix element {\color{black}
as shown in Eq. (\ref{h0})} and the densities of states are given
by equations: 
\begin{equation}
\nu_{j}(\epsilon)=-\frac{1}{\pi}\cosh\left(\frac{\epsilon}{2T}\right)\int_{-\infty}^{\infty}G_{j}\left({\displaystyle \frac{1}{2T}+it}\right)e^{i\epsilon t}dt,\label{eq:DoS}
\end{equation}
where $G_{j}(1/2T+it)$ are exact Green's Functions (GF) in the terminals
$j=1,2$, $f_{1}(\epsilon)=f(\epsilon+e\Delta V/2,T+\Delta T/2)$,
$f_{2}(\epsilon)=f(\epsilon-e\Delta V/2,T-\Delta T/2)$ are corresponding
Fermi distribution functions, and $df(\epsilon)/d\epsilon=-1/[4T\cosh^{2}(\epsilon/2T)]$,
$df(\epsilon)/dT=\epsilon/[4T^{2}\cosh^{2}(\epsilon/2T)]$. The currents
in linear response regime are given by 
\begin{eqnarray}
I_{e} & = & 2\pi e^{2}|t|^{2}\frac{\Delta V}{4T}\int_{-\infty}^{\infty}d\epsilon\frac{\nu_{1}(\epsilon)\nu_{2}(\epsilon)}{\cosh^{2}\left({\displaystyle \frac{\epsilon}{2T}}\right)}-2\pi e|t|^{2}\frac{\Delta T}{4T^{2}}\int_{-\infty}^{\infty}d\epsilon\frac{\epsilon\nu_{1}(\epsilon)\nu_{2}(\epsilon)}{\cosh^{2}\left({\displaystyle \frac{\epsilon}{2T}}\right)},\label{eq:current_e}\\
I_{h} & = & -2\pi e|t|^{2}\frac{\Delta V}{4T}\int_{-\infty}^{\infty}d\epsilon\frac{\epsilon\nu_{1}(\epsilon)\nu_{2}(\epsilon)}{\cosh^{2}\left({\displaystyle \frac{\epsilon}{2T}}\right)}+2\pi|t|^{2}\frac{\Delta T}{4T^{2}}\int_{-\infty}^{\infty}d\epsilon\frac{\epsilon^{2}\nu_{1}(\epsilon)\nu_{2}(\epsilon)}{\cosh^{2}\left({\displaystyle \frac{\epsilon}{2T}}\right)}.\label{eq:current_h}
\end{eqnarray}
Following Onsager's theory \cite{Onsager}, we calculate the thermoelectric
coefficients as follows: 
\begin{equation}
L_{ee}=\left.T\frac{\partial I}{\partial\Delta V}\right|_{\Delta T=0}=\frac{\pi e^{2}|t|^{2}}{2}\int_{-\infty}^{\infty}d\epsilon\frac{\nu_{1}(\epsilon)\nu_{2}(\epsilon)}{\cosh^{2}\left({\displaystyle \frac{\epsilon}{2T}}\right)},
\end{equation}
\begin{equation}
L_{he}=\left.T\frac{\partial I_{h}}{\partial\Delta V}\right|_{\Delta T=0}=-\frac{\pi e|t|^{2}}{2}\int_{-\infty}^{\infty}d\epsilon\frac{\epsilon\nu_{1}(\epsilon)\nu_{2}(\epsilon)}{\cosh^{2}\left({\displaystyle \frac{\epsilon}{2T}}\right)}=L_{eh}=\left.T^{2}\frac{\partial I_{e}}{\partial\Delta T}\right|_{\Delta V=0}\,,
\end{equation}
\begin{equation}
L_{hh}=\left.T^{2}\frac{\partial I_{h}}{\partial\Delta T}\right|_{\Delta V=0}=\frac{\pi|t|^{2}}{2}\int_{-\infty}^{\infty}d\epsilon\frac{\epsilon^{2}\nu_{1}(\epsilon)\nu_{2}(\epsilon)}{\cosh^{2}\left({\displaystyle \frac{\epsilon}{2T}}\right)}.
\end{equation}
As a result, the heat conductance is defined as 
\begin{equation}
G_{H}=\left.\frac{\partial I_{h}}{\partial\Delta T}\right|_{I_{e}=0}=\frac{1}{T^{2}}\left[L_{hh}-\frac{L_{he}^{2}}{L_{ee}}\right].\label{eq:GH}
\end{equation}
Plugging in the densities of states in Eq. (\ref{eq:DoS}) we get
formulas of the thermoelectric coefficients. The last step is to parametrize
the exact GF's at imaginary times as $G_{j}(\tau_{j})$$=$$-\nu_{0j}$$\pi$$T$$\left[\sin\left(\pi T\tau_{j}\right)\right]^{-1}\times$
$K_{j}(\tau_{j})$ with $\nu_{0j}$ is the density of states in the
dot computed in the absence of renormalization effects associated
with electron-electron interaction . All effects of interaction and
scattering are accounted for by the correlator $K_{j}(\tau_{j})=\langle T_{\tau_{j}}\hat{F}_{j}(\tau_{j})\hat{F}_{j}^{\dagger}(0)\rangle$
\cite{furusakimatveevprb,andreevmatveev}. It is convenient to introduce
a notation $G_{C}=2\pi e^{2}\nu_{01}\nu_{02}|t|^{2}$ for the conductance
of the tunnel (central) area between two terminals. %To evaluate the
%integrals, we introduce new variables $t_{1}+t_{2}=2t$ and $t_{1}-t_{2}=2\tau$.
%The Jacobian of the transformation $J=\left|\partial(t_{1},t_{2})/\partial(t,\tau)\right|=2$.
{\color{black} Substituting Eq. (\ref{eq:DoS}) into Eqs. (\ref{eq:current_e}),
(\ref{eq:current_h}) and performing integration over frequency we
obtain after some simplification the general formulas for the thermoelectric
coefficients of the two QD-QPC structure nano-device as}: %%%%%%%%%%%%%%
\begin{eqnarray}
L_{ee}=\frac{\pi}{2}G_{C}T^{2}\int_{-\infty}^{\infty}\frac{d\tau}{\cosh^{2}(\pi T\tau)}{\displaystyle K_{1}\left(\frac{1}{2T}+i\tau\right)K_{2}\left(\frac{1}{2T}-i\tau\right),}\label{eq:general_G}
\end{eqnarray}
\begin{eqnarray}
L_{he}=-\frac{i\pi G_{C}T^{2}}{4e}\int_{-\infty}^{\infty}\frac{d\tau}{\cosh^{2}(\pi T\tau)}\left[\left(\partial_{\tau}K_{1}\left(\frac{1}{2T}+i\tau\right)\right)K_{2}\left(\frac{1}{2T}-i\tau\right)\right.\nonumber \\
-\left.K_{1}\left(\frac{1}{2T}+i\tau\right)\left(\partial_{\tau}K_{2}\left(\frac{1}{2T}-i\tau\right)\right)\right],\label{eq:general_GT}
\end{eqnarray}
\begin{eqnarray}
L_{hh}=\frac{G_{C}}{2e^{2}}\pi T^{2}\int_{-\infty}^{\infty}d\tau\left\{ \frac{\pi^{2}T^{2}\left[2-\cosh^{2}(\pi T\tau)\right]}{\cosh^{4}(\pi T\tau)}K_{1}\left({\displaystyle \frac{1}{2T}+i\tau}\right)K_{2}\left({\displaystyle \frac{1}{2T}-i\tau}\right)\right.\nonumber \\
\left.+\frac{1}{\cosh^{2}(\pi T\tau)}\partial_{\tau}K_{1}\left({\displaystyle \frac{1}{2T}+i\tau}\right)\partial_{\tau}K_{2}\left({\displaystyle \frac{1}{2T}-i\tau}\right)\right\} .\label{eq:general_GH}
\end{eqnarray}
%%%%%%%%%%%%%5
The computation of thermoelectric coefficients in Eqs.(\ref{eq:general_G})-(\ref{eq:general_GH})
essentially needs the explicit form of the electron correlators $K_{j}(1/2T\pm i\tau)$.
It depends on the number of conduction channels corresponding to the
number of QPCs connecting the QD with the electrode of the site $j$.
For our purpose of investigating the physical pictures of the Fermi
liquid and non-Fermi liquid states, it is sufficient to consider the
single channel and two-channel charge Kondo effects. In the next section
we discuss a weak link between a) two Fermi liquids; b) a Fermi liquid
and a non-Fermi liquid; c) two non-Fermi liquids in Fig. \ref{f1}.

\section{\Mr{Main results}}

Using the above general formulas (\ref{eq:general_G}), (\ref{eq:general_GT}),
and (\ref{eq:general_GH}) to compute the heat conductance as defined
in formula (\ref{eq:GH}), we proceed straightforwardly to calculation
of the thermoelectric coefficients of the model introduced in Sec.
II (see Fig. 1). We label and discuss three important limiting cases
one by one in this section. We assume in all calculations that $T\ll\min\left[E_{C,1},E_{C,2}\right]$
\cite{andreevmatveev}.

\subsection{Fermi liquid -- Fermi liquid }

This situation happens when each QD is connected to the electrode
through a QPC. We assume for illustration purposes that either the
$\text{QPC}_{21}$ and $\text{QPC}_{22}$ in Fig.\ref{f1} are turned
off or {\color{black} the differences between the reflection amplitudes
of the QPCs in a QD-QPC structure, namely} $||r_{11}|-|r_{21}||$
and $||r_{12}|-|r_{22}||$, are big enough \cite{thanh2010,thanh2018}.
{\color{black} The channel asymmetry plays very important role:
the RG controls the flow to the stable FL strong coupling fixed point.}
We thus have single channel Kondo (1CK) on each side. The correlator
$K_{j}\left(\tau\right)$ for spinless case at the first order of
{\color{black} the reflection amplitude} $|r_{j}|$ {\color{black}
of the QPC$\!_{j}$} in the perturbative expansion is \cite{andreevmatveev}
\begin{equation}
\!\!K_{j}\left(\tau\right)=\!\left(\frac{\pi^{2}T}{\gamma E_{C,j}}\right)^{2}\!\!\frac{1}{\sin^{2}\left(\pi T\tau\right)}\left[1-2\gamma\xi|r_{j}|\cos\left(2\pi N_{j}\right)+4\pi^{2}\xi\gamma|r_{j}|\frac{T}{E_{C,j}}\sin\left(2\pi N_{j}\right)\cot\left(\pi T\tau\right)\right],\label{eq:Kernel_FL}
\end{equation}
with $\xi=1.59$ is a numerical constant, $\gamma=e^{\boldsymbol{C}},\,\boldsymbol{C}\approx0.5772$
is the Euler's constant \cite{andreevmatveev}. We embed this correlator
into Eqs. (\ref{eq:general_G},\ref{eq:general_GT},\ref{eq:general_GH}),
we obtain formulas for the thermoelectric coefficients as %%%%%%%%%%%%%
\textcolor{black}{
\begin{align}
L_{ee} & =A_{1}^{0}G_{C}\frac{T^{5}}{E_{C,1}^{2}E_{C,2}^{2}}\left[1-2\gamma\xi|r_{1}|\cos\left(2\pi N_{1}\right)-2\gamma\xi|r_{2}|\cos\left(2\pi N_{2}\right)\right],
\end{align}
}with $A_{1}^{0}=8\pi^{8}/15\gamma^{4}$, \textcolor{black}{
\begin{align}
L_{he} & =-\frac{A_{2}^{0}G_{C}}{e}\frac{T^{7}}{E_{C,1}^{2}E_{C,2}^{2}}\left[\frac{|r_{2}|}{E_{C,2}}\sin\left(2\pi N_{2}\right)+\frac{|r_{1}|}{E_{C,1}}\sin\left(2\pi N_{1}\right)\right],
\end{align}
}with $A_{2}^{0}=32\xi\pi^{11}/35\gamma^{3}$, and \textcolor{black}{
\begin{eqnarray}
L_{hh} & = & \frac{A_{3}^{0}G_{C}}{e^{2}}\frac{T^{7}}{E_{C,1}^{2}E_{C,2}^{2}}\left[1-2\gamma\xi|r_{1}|\cos\left(2\pi N_{1}\right)-2\gamma\xi|r_{2}|\cos\left(2\pi N_{2}\right)\right],
\end{eqnarray}
}with $A_{3}^{0}=24\pi^{10}/35\gamma^{4}$. So, the heat conductance
$G_{H}$ at the lowest order of temperature and reflection amplitudes
is \textcolor{black}{
\begin{align}
G_{H} & =\frac{A_{3}^{0}G_{C}}{e^{2}}\frac{T^{5}}{E_{C,1}^{2}E_{C,2}^{2}}\left[1-2\gamma\xi|r_{1}|\cos\left(2\pi N_{1}\right)-2\gamma\xi|r_{2}|\cos\left(2\pi N_{2}\right)\right].\label{eq:GH_FL}
\end{align}
}%%%%%%%%%%%%%%%%
\begin{figure}[t]
\!\!\!\!\!\!\!\! \includegraphics[height=42mm]{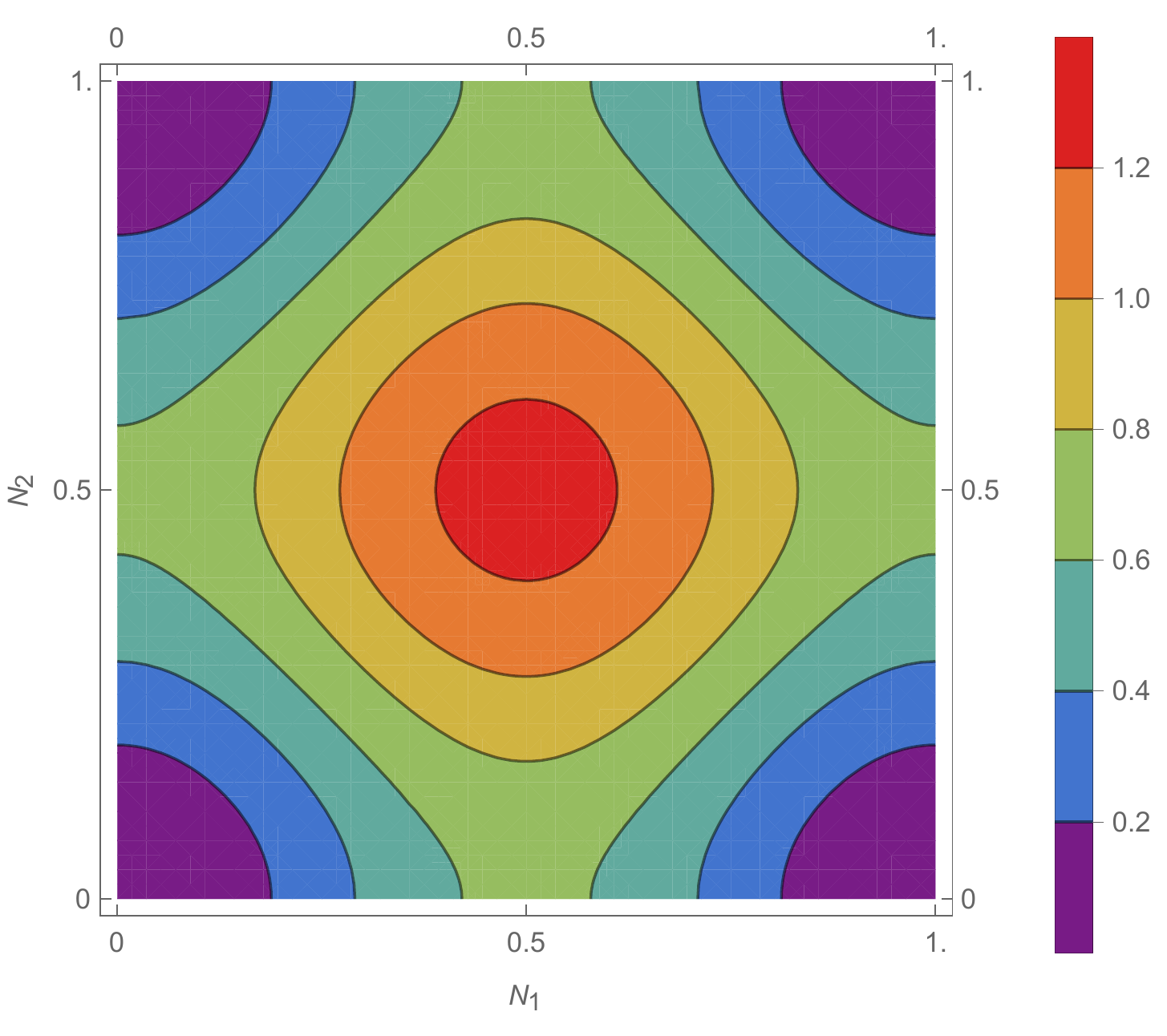}  \includegraphics[height=42mm]{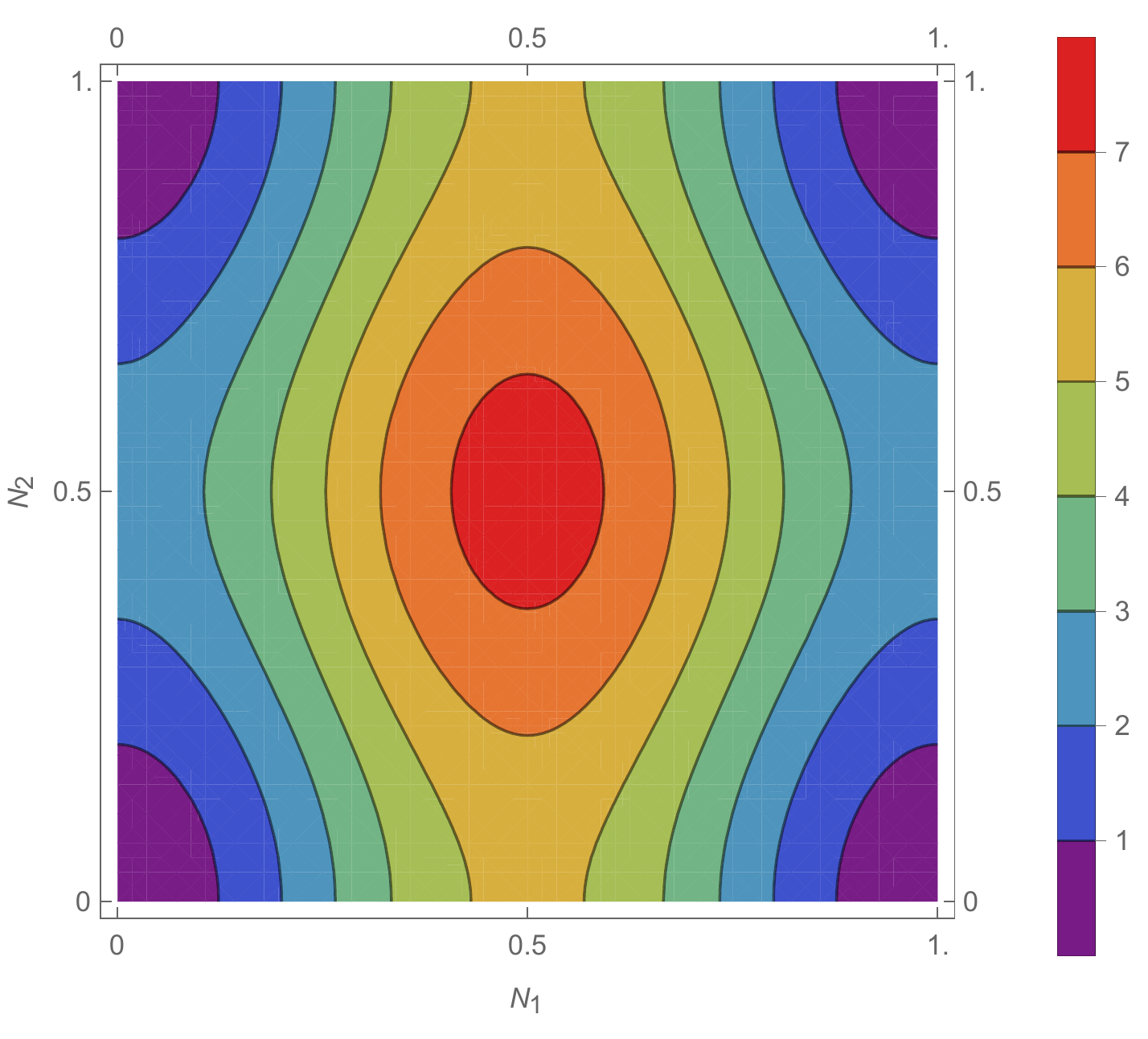}
 \includegraphics[height=42mm]{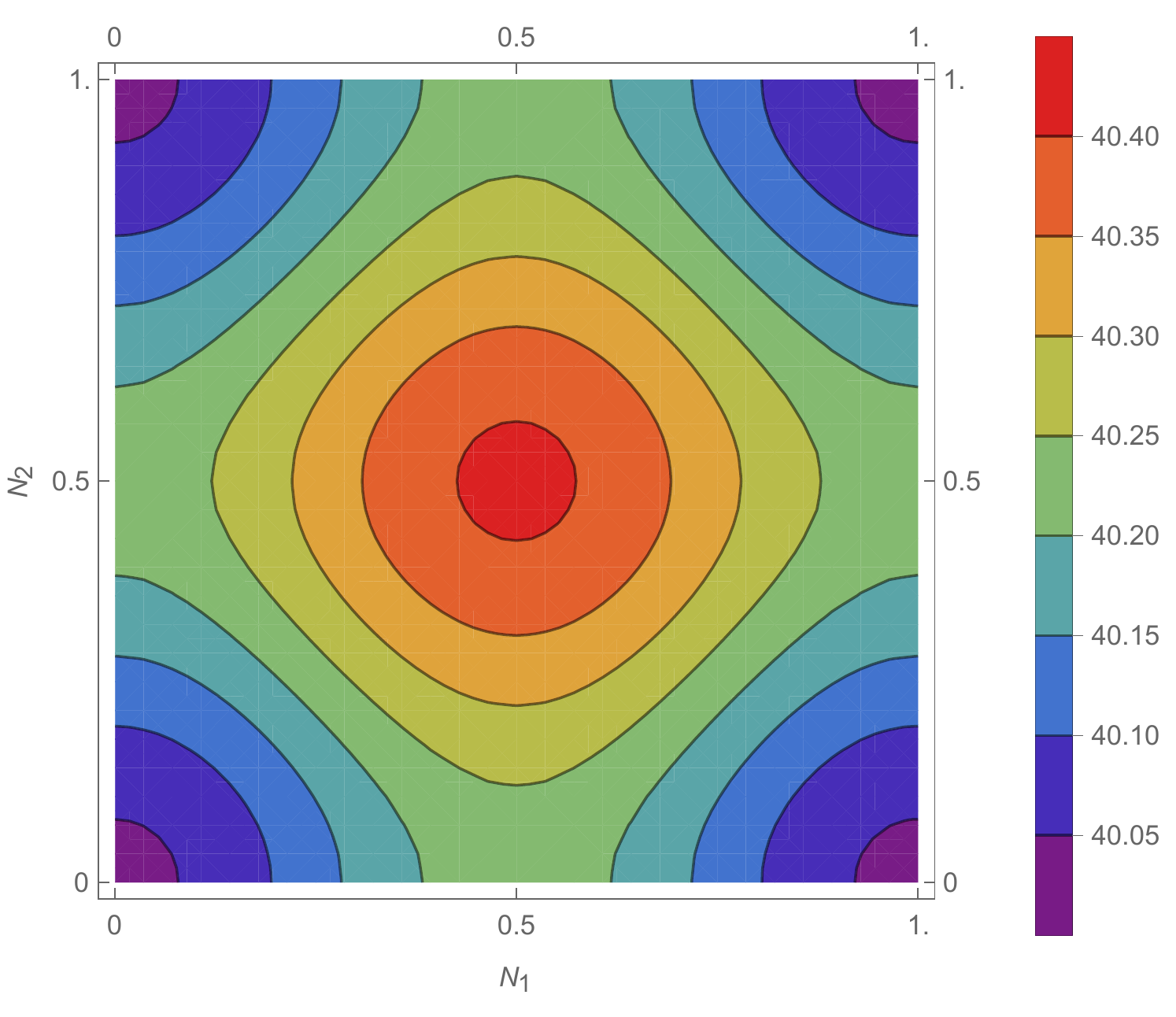}
\caption{(Color online) {\color{black} Contour plots of the heat conductance
$(e^{2}/G_{c})G_{H}\times10^{6}$} as a function of both dimensionless
gate voltages $N_{1}$, $N_{2}$ for the weak link between two Fermi
liquid states (left panel), \textcolor{black}{Fermi liquid state --
non-Fermi liquid state (central panel)}, and two non-Fermi liquid
states (right panel). Here the reflection amplitudes\textcolor{black}{{}
$|r_{1}|=|r_{2}|=0.088$} and temperature $T=0.01$. We choose $E_{C,1}=E_{C,2}=1$.}
\label{f2} 
\end{figure}
The heat conductance as a function of both dimensionless gate voltages
$N_{1}$, $N_{2}$ for the weak link between two Fermi liquid states
is plotted on Fig. \ref{f2} (left panel). We find that the heat conductance
in this case oscillates on both gate voltages $N_{1}$ and $N_{2}$
symmetrically. In order to have the heat conductance being positive,
the values of the reflection amplitudes and temperature must be much
smaller than $E_{C,1},\,E_{C,2}$.

\subsection{Fermi liquid -- Non-Fermi liquid }

This situation happens when the QD in the left side is connected to
the electrode through a QPC (either one QPC is off or $||r_{11}|-|r_{21}||$
is big enough, see explanation in the previous subsection)
while the QD in the right side is connected to the electrode through
two QPCs with the same reflection amplitude $|r_{12}|=|r_{22}|\equiv|r_{2}|$
in Fig. \ref{f1}. We thus have 1CK on the left side and two-channel
Kondo (2CK) on the right side.
The correlator $K_{1}\left(\tau_{1}\right)$ for spinless
case at the first order of $|r_{1}|$ in the perturbative expansion is shown
in Eq. (\ref{eq:Kernel_FL}), while the correlator $K_{2}\left(\tau_{2}\right)$
for spinful case at the second order of $|r_{2}|$ in the perturbative
expansion is \cite{andreevmatveev} \textcolor{black}{
\begin{eqnarray}
K_{2}\left(\tau_{2}\right)&=&\frac{\pi^{2}T}{2\gamma E_{C,2}}\frac{1}{|\sin\left(\pi T\tau_{2}\right)|}\left[1+\frac{2\gamma E_{C2}}{\pi^{4}T}|r_{2}|^{2}\left[\cos\left(2\pi N_{2}\right)+1\right]F\left(\tau_{2}\right)\right.
\nonumber\\
&&\left.-\frac{8\gamma}{\pi^{2}}|r_{2}|^{2}\sin\left(2\pi N_{2}\right)\ln\left[\frac{E_{C,2}}{T}\right]\ln\tan\left(\frac{\pi T\tau_{2}}{2}\right)\right],\label{eq:Kernel_NFL}
\end{eqnarray}
with $F\left(\tau\right)$ expressed in terms of dilogarithm function 
${\rm Li}_2(x)=-\int_0^x t^{-1}\ln(1-t) dt$ \cite{math_func} as
\begin{eqnarray}
F\left(\tau\right) &=& i\left\{ 3\left[-i\pi T\tau\ln\tan\left(\frac{\pi T\tau}{2}\right)-
\text{Li}_2\left[-i\tan\left(\frac{\pi T\tau}{2}\right)\right]+\text{Li}_2\left[i\tan\left(\frac{\pi T\tau}{2}\right)\right]\right]\right.\nonumber\\
 && \left.+i\pi\left(1+T\tau\right)\ln\cot\left(\frac{\pi T\tau}{2}\right)-\text{Li}_2\left[-i\cot\left(\frac{\pi T\tau}{2}\right)\right]+\text{Li}_2\left[i\cot\left(\frac{\pi T\tau}{2}\right)\right]\right\} .
\end{eqnarray}
}Embedding these correlation functions into Eqs. (\ref{eq:general_G})-(\ref{eq:general_GH}), we obtain
%%%%%%%%%%%%%%%
\begin{equation}
L_{ee}=C_{1}^{0}G_{C}\frac{T^{4}}{E_{C,1}^{2}E_{C,2}}\left[1-2\gamma\xi|r_{1}|\cos\left(2\pi N_{1}\right)-\textcolor{black}{C_{1}^{1}|r_{2}|^{2}\frac{E_{C,2}}{T}\cos^{2}\left(\pi N_{2}\right)}\right],
\end{equation}
with $C_{1}^{0}=3\pi^{7}/32\gamma^{3}$ and \textcolor{black}{$C_{1}^{1}=\left(189\zeta\text{\ensuremath{\left(3\right)}}-62\right)16\gamma/27\pi^{5}$,  
$\zeta(s)=\sum_{n=1}^\infty n^{-s}$ 
is the Riemann $\zeta$-function,  
$\zeta\text{\ensuremath{\left(3\right)}}\approx1.2$ is Apery's constant.}
\begin{align}
L_{he} & =-\frac{C_{2}^{0}G_{C}}{e}\frac{T^{5}}{E_{C,1}^{2}E_{C,2}}\left[|r_{1}|\frac{T}{E_{C,1}}\sin\left(2\pi N_{1}\right)+\textcolor{black}{C_{2}^{1}|r_{2}|^{2}\ln\left[\frac{E_{C,2}}{T}\right]\sin\left(2\pi N_{2}\right)}\right],
\end{align}
with $C_{2}^{0}=\pi^{10}\xi/8\gamma^{2}$ and\textcolor{black}{ $C_{2}^{1}=256/25\pi^{5}\xi$},
\begin{equation}
L_{hh}=\frac{C_{3}^{0}G_{C}}{e^{2}}\frac{T^{6}}{E_{C,1}^{2}E_{C,2}}\left[1-2\gamma\xi|r_{1}|\cos\left(2\pi N_{1}\right)-\textcolor{black}{C_{3}^{1}|r_{2}|^{2}\frac{E_{C,2}}{T}\cos^{2}\!\left(\pi N_{2}\right)}\right]
\end{equation}
with $C_{3}^{0}=3\pi^{9}/32\gamma^{3}$ and \textcolor{black}{$C_{3}^{1}=\left(189\zeta\text{\ensuremath{\left(3\right)}}-62\right)16\gamma/27\pi^{5}$}.
So, the heat conductance is 
\begin{equation}
G_{H}=\!\frac{C_{3}^{0}G_{C}}{e^{2}}\frac{T^{4}}{E_{C,1}^{2}E_{C,2}}\!\!\left[\!\!1-2\gamma\xi|r_{1}|\!\cos\left(2\pi N_{1}\!\right)-\textcolor{black}{C_{3}^{1}|r_{2}|^{2}\frac{E_{C,2}}{T}\!\cos^{2}\!\left(\pi N_{2}\!\right)-\frac{\left(C_{2}^{0}\right)^{2}}{C_{3}^{0}C_{1}^{0}}|r_{1}|^{2}\frac{T^{2}}{E_{C,1}^{2}}\!\sin^{2}\!\left(2\pi N_{1}\!\right)}\!\right]\label{eq:GHeat_FLNFL}
\end{equation}
\textcolor{black}{The heat conductance oscillates on the gate voltages $N_{1}$, $N_{2}$
asymmetrically when the Fermi liquid state and non-Fermi liquid state
are weakly coupled as shown in Fig. \ref{f2} (central panel).} The
heat conductance depends on $N_{2}$ [as \textcolor{black}{$|r_{2}|^{2}T^{3}\cos^{2}\left(\pi N_{2}\right)$}]
less strong than on $N_{1}$ [as \textcolor{black}{$|r_{1}|T^{4}\cos\left(2\pi N_{1}\right)$
and $|r_{1}|^{2}T^{6}\sin^{2}\left(2\pi N_{1}\right)$}]. Thus, it
is easy to find that the FL-1CK dominates in the heat conductance.
One may think that the reason is the flow of the heat energy from
the FL-1CK side to the NFL-2CK one but it is not the case. Even though
the NFL-2CK site is at higher temperature, the FL-1CK still contributes
from the first order term. In order to have effects of NFL-2CK considerable
in Eq. (\ref{eq:GHeat_FLNFL}), the reflection amplitude $|r_{2}|$
and temperature must satisfy \textcolor{black}{$T^{-3/2}|r_{2}|/|r_{1}|\geq 49.3718$
and $\left(|r_{2}|^{2}/|r_{1}|\right)/T\geq 9.94$.}
In fact, these
conditions can be satisfied in experiments, we predict that there
exists a temperature $T^{*}$ at which the crossover FL -- NFL happens
in the heat conductance \cite{thanh2018,KK2019}.

\subsection{Non-Fermi liquid -- Non-Fermi liquid }

This regime is realized when both QD in both sides are connected to
the electrode through two QPCs with the same reflection amplitude
$|r_{j}|$, see Fig. \ref{f1}. We thus have 2CK on each side. We
embed the correlator $K_{j}\left(\tau_{j}\right)$ for spinful case
at the second order of $|r_{j}|$ in the perturbative expansion as
in Eq. (\ref{eq:Kernel_NFL}) \cite{andreevmatveev} into Eqs. (\ref{eq:general_G},\ref{eq:general_GT},\ref{eq:general_GH}),
we obtain
\begin{equation}
L_{ee}=P_{1}^{0}G_{C}\frac{T^{3}}{E_{C,1}E_{C,2}}\left[1-\textcolor{black}{P_{1}^{1}\sum_{j=1,2}|r_{j}|^{2}\frac{E_{C,j}}{T}\cos^{2}\left(\pi N_{j}\right)}\right],
\end{equation}
with $P_{1}^{0}=\pi^{4}/6\gamma^{2}$, \textcolor{black}{$P_{1}^{1}=\left(16\ln2-1\right)\gamma/\pi^{3}$,}
\begin{equation}
L_{he}=-\frac{P_{2}^{0}G_{C}}{e}\frac{T^{4}}{E_{C,1}E_{C,2}}\sum_{j=1,2}|r_{j}|^{2}\ln\left[\frac{E_{C,j}}{T}\right]\sin\left(2\pi N_{j}\right),
\end{equation}
with $P_{2}^{0}=3\pi^{4}/16\gamma$,
\begin{eqnarray}
L_{hh} & = & \frac{P_{3}^{0}G_{C}}{e^{2}}\frac{T^{5}}{E_{C,1}E_{C,2}}\left[1-\textcolor{black}{P_{3}^{1}\sum_{j=1,2}|r_{j}|^{2}\frac{E_{C,j}}{T}\cos^{2}\left(\pi N_{j}\right)}\right],
\end{eqnarray}
with $P_{3}^{0}=2\pi^{6}/15\gamma^{2}$, \textcolor{black}{$P_{3}^{1}=\left(16\ln2-49/64\right)\gamma/\pi^{3}$}.
So, the heat conductance \textcolor{black}{at the lowest order which
depends on the gate voltage $N_{j}$ is:}
\begin{eqnarray}
G_{H} & = & \frac{P_{3}^{0}G_{C}}{e^{2}}\frac{T^{3}}{E_{C,1}E_{C,2}}\left[1-\textcolor{black}{P_{3}^{1}\sum_{j=1,2}|r_{j}|^{2}\frac{E_{C,j}}{T}\cos^{2}\left(\pi N_{j}\right)}\right].\label{eq:GHeat_NFL}
\end{eqnarray}
%%%%%%%%%%%%%%
We find that in Eq. (\ref{eq:GHeat_NFL}), following the zero order
in the perturbative series is the\textcolor{black}{{} second} order of
the reflection amplitudes. It means the heat conductance oscillates
weakly on the gate voltage $N_{j}$. The heat conductance as a function
of both dimensionless gate voltages $N_{1}$, $N_{2}$ is plotted
for the weak link between two non-Fermi liquid states on Fig. \ref{f2}
(right panel).

The Eqs. (\ref{eq:GH_FL}), (\ref{eq:GHeat_FLNFL}), and (\ref{eq:GHeat_NFL})
are central results of this paper. The heat conductance is affected
by FL states more strongly than by NFL states. The zero order terms
(the main terms) follow the temperature scaling $T^{5}$, $T^{4}$,
and $T^{3}$ corresponding to the situations coupling between two
FL states, between FL and NFL states, and between two NFL states.
The Coulomb oscillations appear at the first order term when one or
both of the charge Kondo circuits is/are in FL regime while they appear
at the \textcolor{black}{second} order when one or both sides is/are
in NFL states. It is understood that the NFL states originated from
a two-channel charge Kondo with strong fluctuations of the isospin.
Therefore, at the lowest order, the heat conductance \textcolor{black}{is deducted by a percentage which is proportional to $1/T$ (see terms inside the bracket)} concerning NFL states.

\section{\Mr{Conclusion}}

In summary, we have derived equations for the heat conductance of
a quantum circuit containing a weak coupling between two QD-QPC structures
where each one is corresponding to a charge Kondo simulator: either
single channel -- Fermi liquid state or two channel -- non-Fermi
liquid state. The heat conductance is computed in perturbation theory
assuming the smallness of the reflection amplitudes at the QPCs. In
this regime the heat conductance in the complex charge Kondo circuit
is small and consistent with classical explanation for bulk materials
\cite{heatconductance book}. The gate voltage dependence is more
pronounced when the Kondo states are FL than when the Kondo states
are NFL. For the mixed regime when the weak link connects the FL and
NFL states, the FL state dominates in the heat conductance. All regimes
can be experimentally verified in quantum transport measurements with
charge Kondo simulators.

Other interesting directions for future work can be investigating
heat conduction in the presence of Kondo charge correlation in different
complex setups. The extensions of the calculations beyond the linear
response \cite{KK2017} and/or the perturbation theory is a challenging
problem \cite{Karki2022a}. Investigations of the effects of the strong
couplings between quantum dots in the Kondo regime are currently open
for future research \cite{two_islands,Karki2022a}.

\section*{ACKNOWLEDGEMENT}

This research in Hanoi is funded by Vietnam National Foundation for
Science and Technology Development (NAFOSTED) under grant number 103.01-2020.05.
The work of M.K. is conducted within the framework of the Trieste
Institute for Theoretical Quantum Technologies (TQT).

\end{document}